# Cross-platform frequency-domain physical neural networks with identical models and parameters

Zichen Xi[1], Jun Ji[1,*], Liyang Jin[1], Hsuan-Hao Lu[2], Wenjie Xiong[1], and Linbo Shao[1,3,*]
[1] *Bradley Department of Electrical and Computer Engineering, Virginia Tech, Blacksburg, Virginia 24061, USA*
[2] *Elmore Family School of Electrical and Computer Engineering, Purdue University, West Lafayette, Indiana 47907, USA*
[3] *Department of Physics and Center for Quantum Information Science and Engineering (VTQ), Virginia Tech, Blacksburg, Virginia 24061, USA*
[*] *Email: junji@vt.edu; shaolb@vt.edu*

## Abstract
Frequency-domain physical neural networks (PNNs) have emerged as a promising analog computing paradigm, achieving a reduced number of devices, enhanced robustness, and high inference accuracy. However, existing demonstrations across various physical domains, such as optics, acoustics, and electronics, are mostly specific to the target devices or platforms, usually requiring hardware-dependent models and post-fabrication parameter tuning. Here, we demonstrate cross-platform implementations of frequency-domain PNN across different wave-based computing platforms using identical, pre-trained parameters. By exploiting shared second-order nonlinear processes, the PNN model is implemented on optical, microwave, and acoustic-wave platforms without the need of fine tuning the parameters. Evaluated on a unified four-class classification task, the frequency-domain PNN achieves high inference accuracies of 97.6% in optics, 98.4% in electronics, and 98.2% in mechanics. Further, we systematically compare related performance metrics across these three distinct physical domains, paving the way to a universal neural network across different physical platforms.

## Introduction
Artificial neural networks have transformed the landscape of modern technology [1], enabling unprecedented capabilities in computer vision [2-4], natural language processing [5,6], complex scientific computing tasks [7-9], and handling daily life and work tasks. However, as neural network models grow rapidly, their implementation on conventional digital processors, including GPUs, are seeing bottlenecks, such as energy consumption [10,11], processing latency [11,12] and data movement bottlenecks between memory and processing units [13,14]. To address these challenges, wave-based physical neural networks (PNNs) [14-54] have emerged as an attractive computing paradigm. These systems exploit intrinsic wave phenomena [20], such as modulation [22,27] and interference [22,23,27], to efficiently implement the multiply-accumulate (MAC) operations required for matrix-vector multiplication. Photonic PNNs, for example, have been demonstrated using interferometer mesh [22,27,34], electro-optic modulator arrays [28,44] and diffractive optical elements [23,26]. In architectures based on these programmable device arrays [22,27], scaling up computational capacity typically requires additional computing units and control channels. In addition, device-to-device variations due to unintended fabrication variations [31], material dispersions and environmental fluctuations would also cause the computation outputs to deviate from the ideal mathematical model assumed during the neural-network training [42].

Recently, frequency-domain PNN [15,21,35,37,45,46,49,54] provides an alternative approach to high-throughput parallel computing [21,37,45] with more flexible reconfigurability and less susceptibility to device-to-device variations. For example, radiofrequency (RF) photonic processor [45], RF electronic mixer [21], and nonlinear phononic device [37] allow MAC operations to be reconfigured through spectral programming instead of modifying the device configuration. Nonlinear phononic device [37] alleviates the influence of device-to-device variations by encoding input data and neural-network parameters and executing computation on a single device. These demonstrations in photonic, electronic and phononic systems have accelerated tasks such as image and audio classification.

However, existing implementations often incorporate hardware-specific physical responses in their network models [15,45,46,54] or during the training phase [35,49]. Transferring a pre-trained neural-network across different physical domains therefore requires re-training or fine-tuning of the neural-network parameters. This limits the direct deployment of an offline pre-trained network across platforms with different operating characteristics.

In this work, we report a cross-platform frequency-domain PNN framework based on a common mathematical model of second-order spectral convolution. The framework uses identical pre-trained parameters across distinct physical implementations without the need of re-training or parameter fine-tuning. We previously implemented a frequency-domain PNN on our integrated nonlinear phononic platform [37]. Here, we experimentally demonstrate the same two-layer neural-network architecture with two additional distinct implementations: an optical implementation based on a periodically poled lithium niobate (PPLN) module and an electronic implementation based on an RF mixer. All three systems implement the spectral convolution through the second-order nonlinear responses, despite their distinct underlying physical mechanisms. Evaluating using a four-class Fashion-MNIST [55] classification task, we achieve experimental inference accuracies of 97.6% in the optical implementation and 98.4% in the electronic implementation, compared with 98.2% in the phononic implementation. We further compare their nonlinear conversion bandwidths, nonlinear conversion efficiencies, and computing accuracies to identify the common capabilities and platform-dependent characteristics of our frequency-domain PNN.

## Frequency-domain neural network architecture

Our frequency-domain PNNs have two layers (Fig. 1a). Layer 1 is a nonlinear convolution layer and Layer 2 is a fully connected linear mapping layer. In each layer, the input data and corresponding pre-trained parameters [37] are vectorized and concurrently encoded into the frequency spectrum. Each vector element is assigned to a frequency bin; frequency bins within each vector are equally separated by a frequency step $df$. Our second-order nonlinear physical systems perform a spectral self-convolution of the encoded input spectrum in the fundamental frequency band and generate an output spectrum in the second-order frequency band.

Our frequency-domain PNN is benchmarked based on a four-class image classification task using a subset of the Fashion-MNIST dataset containing 28-pixel-by-28-pixel images of T-shirts, trousers, sneakers, and bags. Our two-layer PNN performs inference on each input image as follows:

Layer 1 performs a full convolution of the input image and Layer-1 pre-trained parameters. The input spectrum of Layer 1 consists of 2,000 frequency bins in the fundamental frequency band: 784 bins are encoded with the vectorized pixels of the input 28 × 28 image, and 1,216 bins are encoded with the Layer-1 pre-trained parameters. The output spectrum has 3,999 new frequency bins generated in the second-order frequency band. Among them, the first 2,783 bins are relevant to self-convolution of the input image and the cross-convolution between the input image and pre-trained parameters, and are therefore marked as meaningful outputs of Layer-1 and fed into Layer 2. The remaining 1,216 bins are discarded, as they are solely contributed by the self-convolution of pre-trained parameters.

Layer 2 performs a matrix-vector multiplication between the vector of 2,783 Layer-1 outputs and matrix of 4 × 2,783 Layer-2 pre-trained parameters. The input spectrum of Layer 2 consists of 5 × 2,783 frequency bins in the fundamental frequency band: 4 × 2,783 bins are encoded with the Layer-2 pre-trained parameters, and 2,783 bins are encoded with the Layer-1 outputs. In the output spectrum, the amplitudes at four designated frequency bins represent the resulting matrix-vector product and indicate the class scores. The class with the highest score is identified as the predicted class. Encoding and decoding details are described in Supplementary Note 1.

Our two-layer frequency-domain PNN framework is demonstrated by an optical implementation based on a PPLN module (Fig. 1b) and an electronic implementation based on a megahertz-frequency RF mixer (Fig. 1c). Our previously demonstrated gigahertz-frequency surface-acoustic-wave (SAW) device [37] provides a phononic implementation of the same framework (Fig. 1d). Although these three platforms work at distinct frequency regimes from MHz to THz, they all leverage the same second-order nonlinearity to physically perform the spectral convolution, which is the core mathematical operation of our frequency-domain PNN framework.

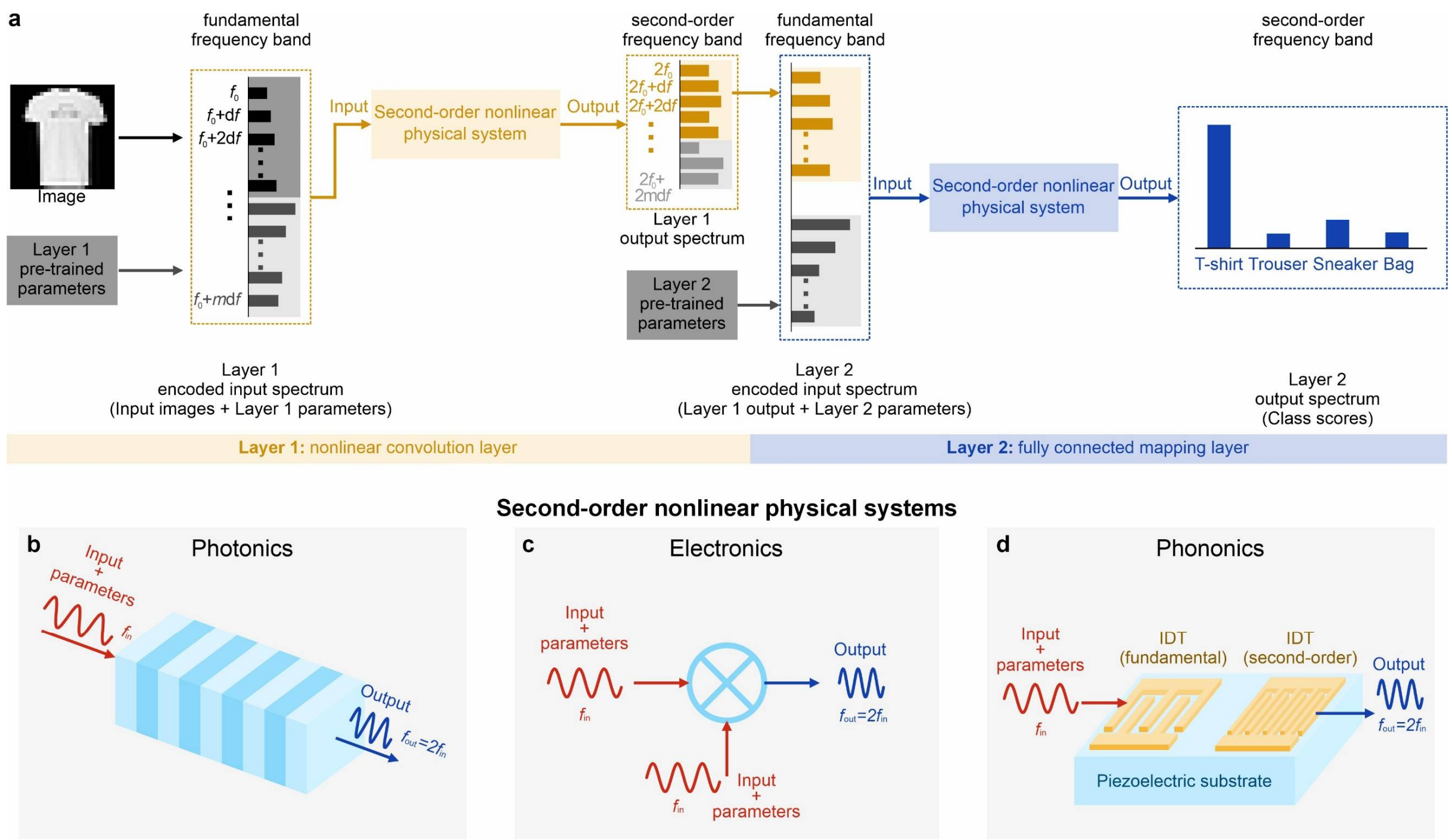

**Figure 1. Introduction to fine-tuning-free frequency-domain neural network using second-order nonlinear physical systems. a.** Diagram of the frequency-domain neural network architecture. The frequency-domain neural network consists of a nonlinear convolution layer (Layer 1) and a fully connected linear mapping layer (Layer 2). Input features and pre-trained parameters are concurrently encoded in the frequency domain at different frequencies in the fundamental frequency band. Output spectra from the second-order nonlinear physical system are read out from the second-order frequency band. The class scores are read out from the Layer-2 output spectrum. **b, c, d.** Representative second-order nonlinear physical systems based on **b.** photonics, **c.** electronics, and **d.** phononics. The photonic system uses second-order nonlinear conversion in a periodically poled lithium niobate (PPLN) waveguide, the electronic system uses a radiofrequency mixer, and the phononic system uses a surface-acoustic-wave device with a pair of interdigital transducers, one for exciting acoustic waves in the fundamental frequency band and the other for receiving acoustic waves in the second-order frequency band.

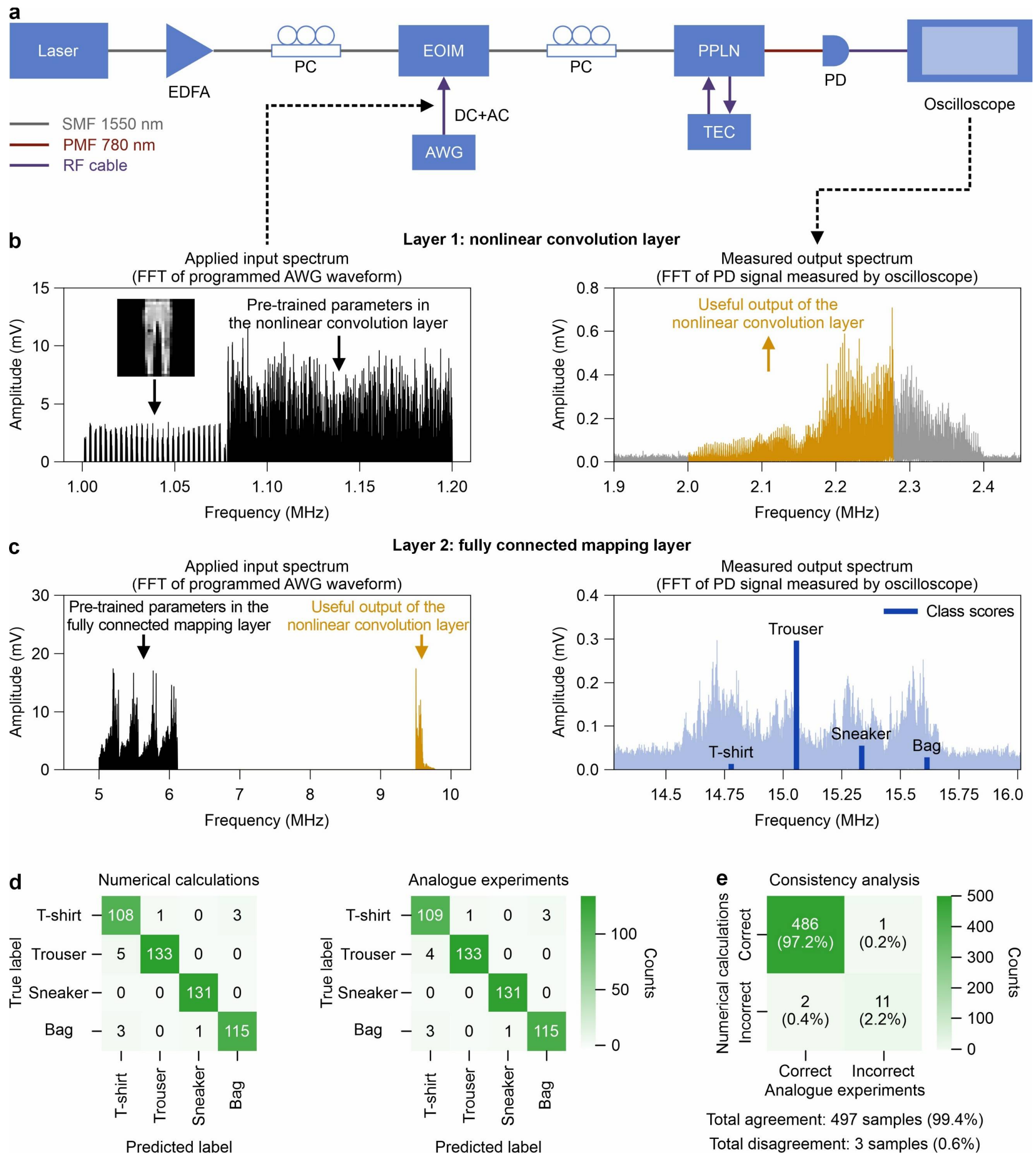

**Figure 2. Optical implementation of frequency-domain neural network using periodically poled lithium niobate (PPLN) for four-class classification. a.** Experimental setup of the optical neural network system. A tunable 1550-nm laser is amplified by an erbium-doped fiber amplifier (EDFA), polarization controlled and then modulated by an electro-optic intensity modulator (EOIM) driven by an arbitrary waveform generator (AWG). The encoded optical signal is polarization-controlled and input to a PPLN module. The generated 780-nm light is detected by a photodetector (PD) and recorded by an oscilloscope. The PPLN module temperature is stabilized by a thermoelectric cooler (TEC). **b.** Spectrum of the nonlinear convolution layer (Layer 1). Left: applied input spectrum calculated as the fast Fourier transform (FFT) of the programmed AWG waveform, containing the encoded image and the pre-trained parameters. Right: measured output spectrum calculated as the FFT of the PD signal measured by the oscilloscope. The orange region indicates the useful frequency bins that contain contributions from the encoded input image. **c.** Spectrum of the fully connected mapping layer (Layer 2). Left: applied input spectrum, containing the pre-trained parameters and the useful output from Layer 1. Right: measured output spectrum. The class scores are read out from the amplitudes of four readout frequency bins. **d.** Confusion matrices of digital calculations and analogue optical experiments over 500 Fashion-MNIST test samples. The numerical calculations account for the EOIM transfer function with nonlinear response. **e.** Consistency analysis between numerical calculations and analogue optical experiments, showing a total agreement of 497 samples (99.4%) and disagreement of 3 samples (0.6%).

## Optical implementation based on a periodically poled lithium niobate module

Figure 2a shows the experimental setup for our optical implementation of the frequency-domain PNN. A continuous-wave tunable laser (Santec TSL-770) operating near 1550 nm provides the optical carrier of 0 dBm, which is amplified by an erbium-doped fiber amplifier (EDFA; Amonics AEDFA-33-B) with gain of 17 dB. A polarization controller adjusts the polarization of the amplified light to match the preferred input polarization of the electro-optic intensity modulator (EOIM; Lucent Technologies 2623NA). The EOIM, biased at its quadrature point, modulates the optical carrier with a baseband signal representing the input features and pre-trained parameters. An arbitrary waveform generator (AWG; Rigol DG922 Pro) provides the DC bias voltage and synthesizes the AC modulating signal for the EOIM. The modulated light is then adjusted to a preferred polarization by a second polarization controller, before entering the fiber-coupled PPLN module (ADvR) for the type-0 second-harmonic generation (SHG). The PPLN module generates the second-harmonic light near 775 nm from the modulated input near 1550 nm. To maintain the phase-matching condition for SHG, the operating temperature of the PPLN module is stabilized by a thermoelectric cooler (TEC). A photodetector (PD) (Newport 1811) measures the generated second-harmonic light, and an oscilloscope (Rohde & Schwarz MXO44) records the resulting electrical waveform for subsequent spectral analysis.

We use one image as an example to illustrate the encoding and decoding procedures of Layer 1 and Layer 2. Figures 2b and 2c (left side) show the numerically generated input spectra used to modulate the carrier light, while Figs. 2b and 2c (right side) show the experimentally measured electrical output spectra from the PD. Photodetection converts the intensity modulation of the optical signal into an electrical signal at MHz frequencies.

In Layer 1, the 784 grayscale pixel values are encoded to frequency bins spanning from 1.0001 to 1.0784 MHz in the baseband, with a frequency step $df$ of 100 Hz. The amplitude at each bin is proportional to the corresponding pixel value. The 1,216 pre-trained parameters are encoded in the following bins from 1.0785 to 1.2000 MHz, with the same $df = 100$ Hz. For clarity, the plotted spectra show magnitudes only and therefore do not reveal the signs of these parameters. For decoding, the 2,783 image-related output bins from 2.0002 to 2.2784 MHz (highlighted in orange in Fig. 2b, right) are read out as the Layer-1 outputs. These outputs are then used to regenerate input spectrum for Layer 2.

In Layer 2, the 4 × 2,783 pre-trained parameters are encoded from 5.0001 to 6.1132 MHz, and the 2,783 input features from Layer-1 outputs are encoded from 9.5001 to 9.7783 MHz, with a frequency step $df$ of 100 Hz (Fig. 2c, left). The additional frequency separation between the input features and parameters is required to separate their cross-convolution from their self-convolutions in the second-order frequency band. The cross-convolution in the output spectrum spans from 14.5002 to 15.8915 MHz. For decoding, the amplitudes at four designated frequency bins (highlighted by the blue thick bars) within this band are read out as the class scores (Fig. 2c, right). The class with the highest score is selected as the predicted class. More encoding and decoding details are provided in Supplementary Note 1 and Supplementary Table 1.

The optical implementation achieves inference accuracies of 97.4% and 97.6% in the numerical calculations and analogue experiments, respectively (Fig. 2d). The numerical calculations and analogue experiments agree on 497 of 500 test images (99.4%; Fig. 2e). These results demonstrate that our optical implementation closely reproduces the predictions of the numerical model.

We further numerically investigate the effect of modulation nonlinearity on the inference accuracy. At large modulation depth $\beta = \pi V_{\mathrm{pp}} / V_\pi$, the sinusoidal response of EOIM will distort the modulation from a linear response [56]. Here, $V_{\mathrm{pp}}$ is the peak-to-peak voltage of the AC signal applied to the EOIM and $V_\pi$ is the half-wave voltage of EOIM, measured to be 3.46 V for our case (Supplementary Figure 1). In numerical calculations, we swept the modulation depth $\beta$ from $0.1\pi$ to $1.9\pi$ in both Layer 1 and Layer 2 (Supplementary Figure 2). With a small $\beta$ of $0.1\pi$ in both layers, the calculated inference accuracy of the optical implementation reaches 98.4%, matching that of the numerical model [37]. With the Layer-2 $\beta$ fixed at $0.1\pi$, increasing the Layer-1 $\beta$ from $0.1\pi$ to $1.9\pi$ reduces the inference accuracy from 98.4% to 55.4%. In contrast, when the Layer-1 $\beta$ is fixed at $0.1\pi$, the same variation in Layer-2 $\beta$ changes the inference accuracy from 98.4% to 96.2%. Detailed results are shown in Supplementary Figure 2. This indicates that different layers may have different tolerance to the additional device nonlinearity. In practice, although reducing $V_{\mathrm{pp}}$ improves the modulation linearity, it lowers the measured spectral signal-to-noise ratio (SNR) in the experiments (Supplementary Figure 3). In the experiments shown in Fig. 2, we trade off the modulation linearity and spectral SNR by setting $\beta$ to be $0.72\pi$ and $1.45\pi$ (corresponding to $V_{\mathrm{pp}}$ = 2.5 V and 5.0 V) for Layers 1 and 2, respectively.

## Electronic implementation based on an RF mixer

Figure 3a shows the experimental setup for our electronic implementation of the frequency-domain PNN. The AWG synthesizes a time-domain waveform whose frequency spectrum is encoded with the input features and pre-trained parameters. The waveform is then split into two nominally identical branches via an SMA-T connector and fed to the two input ports of a RF mixer (Analog Device AD834). The RF mixer multiplies two inputs, producing an output proportional to the square of the input waveform. This quadratic response physically implements self-convolution of the encoded input spectrum in the frequency domain. The output waveform is then recorded by an oscilloscope.

Figures 3b and 3c illustrate the two-layer inference process for the same test image used in Figs. 2b and 2c. The electronic implementation used the same pre-trained parameters, encoding, and decoding procedures as the optical implementation, with different encoding and decoding frequencies. In Layer 1, the input spectrum spans from 10.0005 to 11.0000 MHz. In Layer 2, the pre-trained parameters and Layer-1 outputs occupy 15.0005 to 20.5660 MHz and 28.0005 to 29.3915 MHz, respectively. A frequency step $df$ of 500 Hz is used in both layers. Four designated frequency bins within the cross-convolution band from 43.0010 to 49.9575 MHz are read out as the class scores (Fig. 3c, right). The use of different frequency bins in the electronic and optical implementations demonstrates that the PNN architecture can accommodate frequency assignments tailored to different physical platforms. The experimental encoding settings for electronic implementation are summarized in Supplementary Table 2.

The electronic implementation achieves inference accuracies of 98.4% in both numerical calculations and analogue experiments (Fig. 3d), which are slightly higher than the optical implementation (numerical: 97.4%; analogue: 97.6%). The difference is likely attributed to the nonlinear modulation distortion introduced by the EOIM in the optical implementation, as discussed in the previous section. The consistency analysis shows agreement of 100 % between numerical calculations and analogue experiments (Fig. 3e).

## Platform performance comparison

In Fig. 4, we compare three key performance metrics across the optical, electronic and phononic implementations: (1) nonlinear conversion bandwidth, (2) nonlinear conversion efficiency, and (3) calculation accuracy quantified by the normalized mean squared error (NMSE). We note that the results shown in Figs. 4c, 4f, and 4i are replots based on data of our previous work [37].

We measured the frequency-dependent second-harmonic generation (SHG) spectra across the three platforms, with the shaded regions indicating the nonlinear conversion bandwidth (Figs. 4a to 4c). The optical PPLN module exhibits a nonlinear conversion bandwidth of 25.9 GHz (Fig. 4a). The RF mixer features two operating bands with bandwidths of 4.5 MHz ($BW_1$) and 20 MHz ($BW_2$) (Fig. 4b). The $BW_2$ is used for the experiments in Fig. 3. The integrated phononic device provides a nonlinear conversion bandwidth of 0.6 MHz (Fig. 4c).

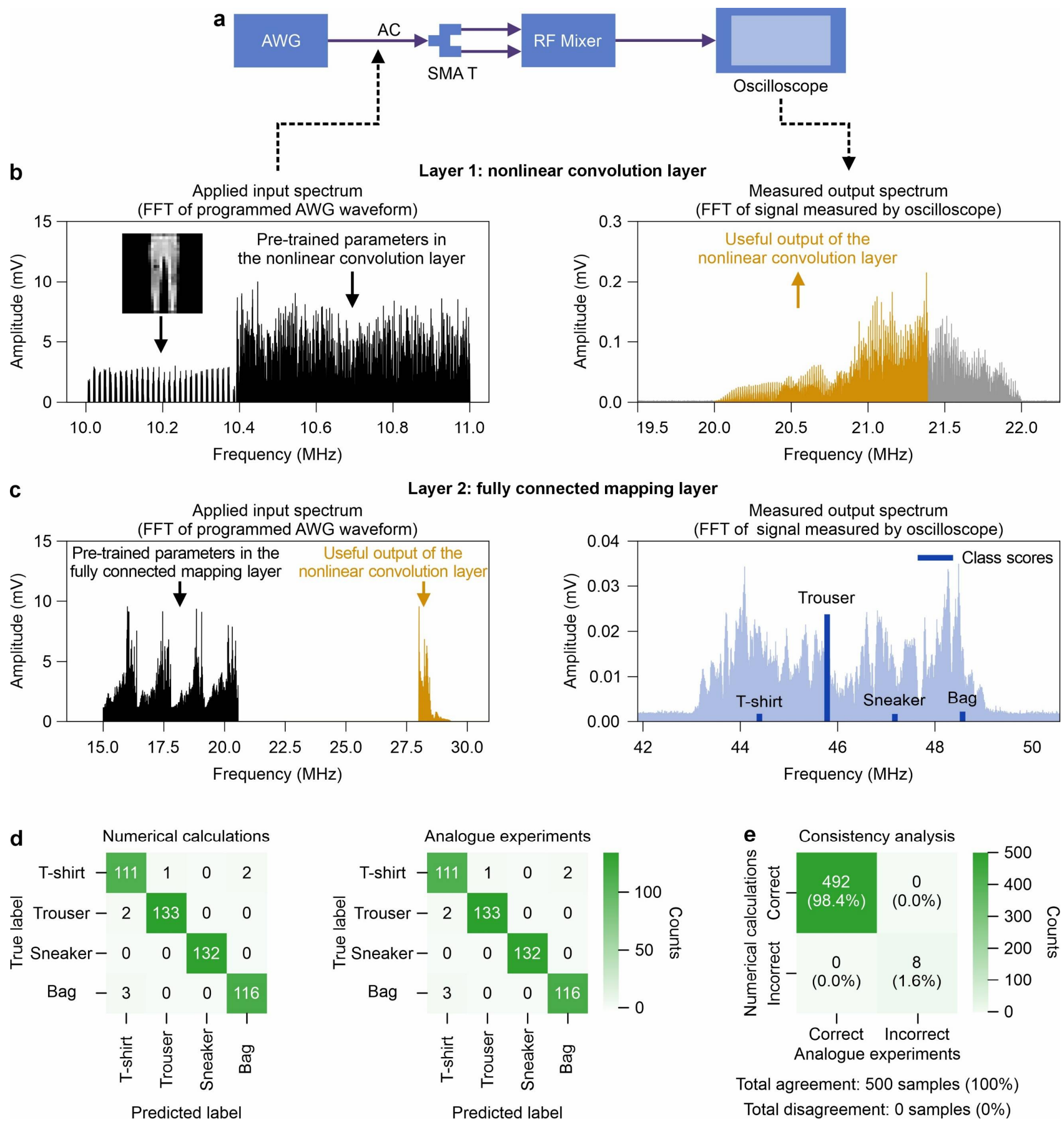

**Figure 3. Electronic implementation of the frequency-domain neural network using a radiofrequency (RF) mixer for four-class classification. a.** Experimental setup of the electronic neural network system. The programmed waveform from arbitrary waveform generator (AWG) is split by an SMA T connector and input to the RF mixer. The mixer output is recorded by an oscilloscope for further spectral analysis. **b.** Spectrum of the nonlinear convolution layer (Layer 1). Left: applied input spectrum calculated as the FFT of the programmed AWG waveform, containing the encoded image and the pre-trained parameters. Right: measured output spectrum calculated as the FFT of the mixer output measured by the oscilloscope. The orange region indicates the useful frequency bins that contain contributions from the encoded input image. **c.** Spectrum of the fully connected mapping layer (Layer 2). Left: reconstructed input spectrum, containing the pre-trained parameters and the useful output from Layer 1. Right: measured output spectrum. The class scores are read out from the amplitudes of four readout frequency bins. **d.** Confusion matrices of numerical calculations and analogue electronic experiments over 500 Fashion-MNIST test samples. **e.** Consistency analysis between numerical calculations and analogue electronic experiments, showing a total agreement of 500 samples (100%).

A larger nonlinear conversion bandwidth can provide wider spectral range for frequency encoding. For a fixed frequency spacing $df$, a wider range can accommodate more encoded frequency bins, enabling more efficient parallel computing. Alternatively, for a fixed number of encoded frequency bins, it can allow a larger $df$. As a result, the minimum signal acquisition time $T = 1/df$ required to resolve adjacent bins can be shorter, resulting in a lower processing latency and a higher computational throughput. The substantially larger nonlinear conversion bandwidth of the optical platform therefore offers greater flexibility than electronic and phononic platforms in balancing computational dimensionality, latency, and throughput.

However, variations in SHG output across the nonlinear conversion bandwidth could introduce unintended weighting into the spectral convolution, potentially reducing inference accuracy. In the optical implementation, the encoded frequency bins span only several megahertz, which is negligible compared to the 25.9-GHz nonlinear conversion bandwidth. Therefore, the SHG output is expected to remain uniform within the encoded frequency range, and minimal weighting is introduced. In comparison, the Layer-1 encoding spans approximately 5% and 33% of the nonlinear conversion bandwidth in the electronic and phononic implementations, respectively, while the corresponding Layer-2 encoding spans approximately 72% and 100%. Although these encodings occupy a larger fraction of the nonlinear conversion bandwidth, both implementations nevertheless maintain high inference accuracies. This robustness suggests that the pre-trained parameters can tolerate moderate nonuniformity in the SHG output across the encoded frequency range.

To quantify the nonlinear conversion efficiency, we measured the SHG output power $P_{\mathrm{SHG}}$ as a function of the input power $P_{\mathrm{in}}$ for the three platforms (Figs. 4d to 4f). The SHG output power exhibits an approximately quadratic dependence on the input power, with fitted slopes on a log-log scale of 2.01, 1.99 and 1.96 for the optical, electronic, and phononic platforms, respectively. From these measurements, we evaluate the nonlinear conversion efficiency as $\eta = P_{\mathrm{SHG}}/P_{\mathrm{in}}^2$. This efficiency is evaluated at the system level, including the input and output coupling interfaces. The measured nonlinear conversion efficiencies are 2.5% $W^{-1}$, 7.1% $W^{-1}$, and 0.063% $W^{-1}$ for the optical, electronic, and phononic implementations, respectively. The corresponding measurement configurations are provided in Supplementary Figure 4. A higher nonlinear conversion efficiency reduces the input power required to produce a given SHG output power, resulting in lower energy consumption of carrier and/or encoding signals.

We also quantified the computing accuracy of the three implementations using the NMSE between the experimentally measured and numerically calculated outputs (Figs. 4g to 4i). Each point represents the NMSE for one of the 500 test samples, with the solid lines indicating the average values. For all three platforms, Layer 1 exhibits relatively low and tightly distributed NMSE, whereas larger NMSE values are observed in Layer 2. The increased error could arise from the accumulation and propagation of physical imperfections through the cascaded computation. We note that the NMSE values of the two layers are obtained from output vectors of substantially different dimensions, with 2,783 bins in Layer 1 and four class scores in Layer 2. Despite the different bandwidths and nonlinear conversion efficiencies, all implementations reproduce the inferences of the common numerical model with high fidelity.

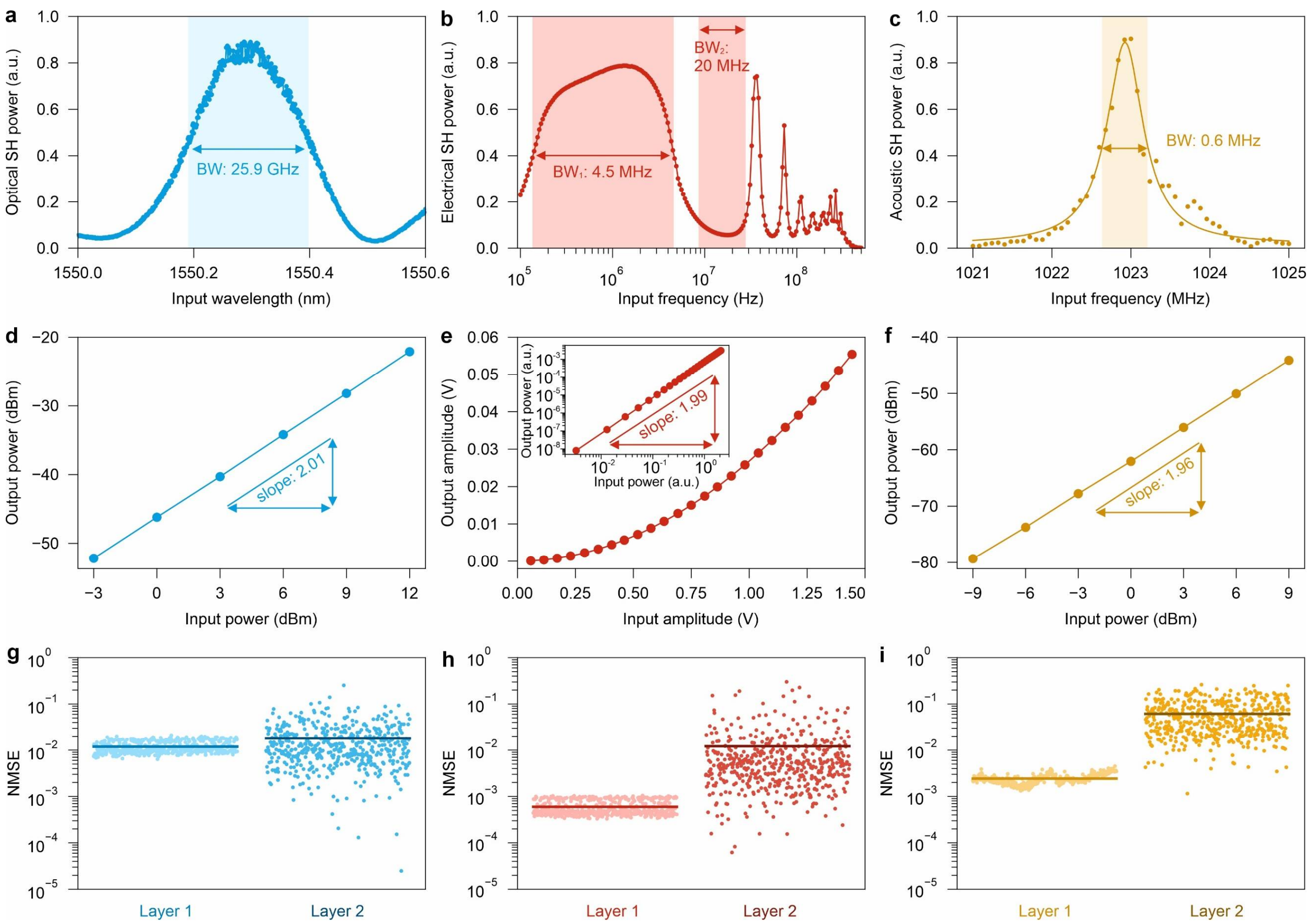

**Figure 4. Physical characteristics across different platforms used for frequency-domain neural network computing. a, b, c.** Normalized frequency-dependent second-harmonic generation (SHG) in **a.** the optical periodically poled lithium niobate (PPLN) module, **b.** the electronic radiofrequency (RF) mixer module, and **c.** the integrated phononic device. The shaded regions indicate the nonlinear conversion bandwidths that are potentially usable for frequency encoding. **d, e, f.** Power-dependent SHG in **d.** the optical, **e.** the electronic, and **f.** the phononic implementations. The extracted slopes near 2 confirm the quadratic dependence of the output power on the input power. **g, h, i.** Normalized mean squared error (NMSE) between measured analogue outputs and numerically calculated results for the same four-class classification inference task shown in Figs. 2 and 3, using **g.** optical, **h.** electronic, and **i.** phononic implementations. The solid lines represent the average normalized mean squared error (NMSE) of 500 test images. All phononic data are adapted from our previous work to provide broader cross-platform context. Data in **c**, **f**, and **i** are from our previous work [37].

## Conclusion and outlook

Overall, we demonstrate a frequency-domain PNN that can be implemented across substantially different physical platforms using the same pre-trained neural-network parameters without platform-specific fine-tuning. The platform independence demonstrated here does not require identical carrier frequency, encoding frequencies, nonlinear conversion efficiency, physical interface, and driving condition. These results establish a basis for deploying pre-trained neural networks across different analogue physical systems. More broadly, the underlying concept is not restricted to the three systems investigated here. Other physical platforms that provide controllable second-order nonlinear interactions could implement the same frequency-domain computing framework. Together, these results suggest a pathway toward hardware-independent analogue neural networks in which a pre-trained model can be deployed across heterogeneous physical computing platforms while the hardware implementation is selected according to the requirements of a particular application.

## Acknowledgments

Research was partially supported by the Air Force Office of Scientific Research (AFOSR) and was accomplished under Grant Number W911NF-23-1-0235. Z.X. was in part supported by the graduate internship in the Quantum Information Science Section of the Computational Sciences and Engineering Division at Oak Ridge National Laboratory (ORNL). This internship was supported by ORNL’s Laboratory Directed Research and Development Program. The views and conclusions contained in this document are those of the authors and do not necessarily reflect the position or the policy of the Government. No official endorsement should be inferred. Approved for public release; distribution is unlimited.

## Author contribution

Z.X performed the experiments with help from L.S., L.J., and J.J.. L.S., W.X., and H.-H.L. designed the project. Z.X. and J.J. drafted the manuscript with revisions from all other authors. L.S. supervised the project.

## Conflict of interests

The authors have no conflicts to disclose.

## Data availability

The data supports the findings of this study are present in the paper and/or the Supplementary Materials. Additional data that supports the fundings of this study are available from the corresponding author upon reasonable request.

## Supplementary Information

### Supplementary Note 1
### Encoding and decoding methods of frequency-domain PNN framework

Encoding methods:

In the first layer, the input image and the Layer 1 pre-trained parameters are first vectorized into $\boldsymbol{a}$ and $\boldsymbol{b}$, respectively. Vector $\boldsymbol{a} = (a_1, a_2, \dots, a_M)$ has $M$ elements, with each element corresponding to the scaled grayscale value of one pixel of the input image, while the vector $\boldsymbol{b} = (b_1, b_2, \dots, b_N)$ has $N$ elements, with each element corresponding to one scaled pre-trained parameter. These elements are then encoded in the frequency domain as follows:

$$S_{in,1}(\Omega) = \sum_{i=1}^{M} a_i \delta(\Omega - \Omega_i) + \sum_{j=1}^{N} b_j \delta(\Omega - \Omega_j), \tag{S1}$$

where $\Omega_i = \Omega_0 + i\Omega_m, i = 1,2, \dots, M,$ and $\Omega_j = \Omega_0 + (M + j)\Omega_m, j = 1,2, \dots, N$ are angular frequencies assigned to elements of $\boldsymbol{a}$ and $\boldsymbol{b}$, respectively. The condition $\Omega_0 \geq (M + N + 1)\Omega_m$ is required to avoid spectral overlap between the encoded input bins near $\Omega_0$ and the second-order generated bins near $2\Omega_0$ in the encoded-frequency spectrum. $\delta(\cdot)$ denotes the Dirac delta function.

In the second layer, the Layer 2 pre-trained parameters and the useful outputs from Layer 1 are vectorized into $\boldsymbol{c}$ and $\boldsymbol{d}$, respectively. Vector $\boldsymbol{c} = (c_1, c_2, \dots, c_P)$ has $P$ elements, and vector $\boldsymbol{d} = \left(d_1, d_2, \dots, d_Q\right)$ has $Q$ elements, with $P = nQ$, where $n$ denotes the number of classes for a classification task. In our four-class classification task, $n = 4$. Then, vector $\boldsymbol{c}$ and $\boldsymbol{d}$ are encoded into frequency domain as follows:

$$S_{in,2}(\Omega) = \sum_{i=1}^{P} c_i \delta(\Omega - \Omega_i) + \sum_{j=1}^{Q} d_j \delta(\Omega - \Omega_j), \tag{S2}$$

where $\Omega_i = \Omega_0 + i\Omega_m, i = 1,2, \dots, P$, and $\Omega_j = \Omega_0 + (K + Q - j + 1)\Omega_m, j = 1,2, \dots, Q$ are angular frequencies assigned to elements of $\boldsymbol{c}$ and $\boldsymbol{d}$, respectively. The condition $K \geq 2P$ is required to separate the cross-convolution terms between $\boldsymbol{c}$ and $\boldsymbol{d}$ from the self-convolution terms of $\boldsymbol{c}$ and $\boldsymbol{d}$ in the encoded-frequency spectrum. The condition $\Omega_0 \geq (K + Q + 1)\Omega_m$ is required to avoid spectral overlap between the encoded input bins near $\Omega_0$ and the second-order generated bins near $2\Omega_0$ in the encoded-frequency spectrum.

In physical implementations, the encoded spectra given in Eqs. (S1) and (S2) are realized by synthesizing time-domain waveforms using the arbitrary waveform generator (AWG). The relative spectral amplitudes of the frequency bins are determined by the encoded input features and pre-trained parameters, while the overall voltage scale is set by the peak-to-peak voltage $V_{\mathrm{pp}}$ of the AWG. The absolute spectral amplitude of each frequency bin is determined by the $V_{\mathrm{pp}}$ value set in the AWG. For example, writing Eq. (S1) in the time domain:

$$s_{in,1}(t) = \sum_{i=1}^{M} a_i \cos(\Omega_i t) + \sum_{j=1}^{N} b_j \cos(\Omega_j t). \tag{S3}$$

The corresponding waveform programmed into the AWG is

$$s_{AWG,1}(t) = \frac{V_{\mathrm{pp}}}{2} \cdot \frac{s_{in,1}(t)}{\max\left|s_{in,1}(t)\right|}, \tag{S4}$$

where $\max\left|s_{in,1}(t)\right|$ denotes the maximum absolute value of the waveform over the programmed time window.

Decoding methods:

In Layer 1, the decoding frequencies are determined directly from the frequency assignments in Eq. (S1). In the second-order frequency band around $2\Omega_0$, the sum-frequency bins generated by $a-a$, $a-b$, $b-b$ interactions occur at the following frequencies:

$$\Omega_{aa} = 2\Omega_0 + (i + i')\Omega_m,\ \ i, i' = 1,2, \dots, M,$$
$$\Omega_{ab} = 2\Omega_0 + (M + i + j)\Omega_m,\ \ i = 1,2, \dots, M,\ \ j = 1,2, \dots, N,$$
$$\Omega_{bb} = 2\Omega_0 + (2M + j + j')\Omega_m,\ \ j, j' = 1,2, \dots, N.$$

The bins generated by the $a-a$ and $a-b$ interactions contain contributions from the input image vector $\boldsymbol{a}$, whereas the $b-b$ interaction depends only on the Layer-1 pre-trained parameter vector $\boldsymbol{b}$. Therefore, the Layer-1 output is decoded at the discrete frequencies:

$$\Omega_k = 2\Omega_0 + (k + 1)\Omega_m,\ \ k = 1,2, \dots, 2M + N - 1.$$

These decoding frequencies span from $2\Omega_0 + 2\Omega_m$ to $2\Omega_0 + (2M + N)\Omega_m$, with a frequency spacing of $\Omega_m$. The remaining $N$ higher-frequency bins, from $2\Omega_0 + (2M + N + 1)\Omega_m$ to $2\Omega_0 + (2M + 2N)\Omega_m$, contain only contributions from the $b-b$ interaction are therefore discarded. The amplitudes of the $(2M + N - 1)$ decoded frequency bins are squared elementwise to form the Layer-2 input vector $\boldsymbol{d}$ according to Eq. (S2).

In Layer 2, because $P = nQ$, the Layer 2 pre-trained parameter vector $\boldsymbol{c}$ is divided into $n$ consecutive blocks of $Q$ elements. The $k$th block corresponds to the weight vector of the $k$th output class:

$$\boldsymbol{c}^{(k)} = \left(c_{kQ+1}, c_{kQ+2}, \dots, c_{(k+1)Q}\right),\ \ k = 0,1, \dots, n - 1.$$

According to the frequency assignments in Eq. (S2), the $j$th element of the $k$th class block and the corresponding element $d_j$ are encoded at:

$$\Omega_{c,k,j} = \Omega_0 + (kQ + j)\Omega_m,\ \ k = 0,1, \dots, n - 1,\ \ j = 1,2, \dots, Q,$$
$$\Omega_{d,j} = \Omega_0 + (K + Q - j + 1)\Omega_m,\ \ j = 1,2, \dots, Q.$$

The sum frequency generated by this corresponding pair is therefore:

$$\Omega_{c,k,j} + \Omega_{d,j} = 2\Omega_0 + [K + (k + 1)Q + 1]\Omega_m \equiv \Omega_{w,k}.$$

The index $j$ cancels from the resulting frequency. Consequently, for a fixed $k$, all $Q$ products between corresponding elements of $\boldsymbol{c}^{(k)}$ and $\boldsymbol{d}$ are generated at the same readout frequency $\Omega_{w,k}$ and are accumulated in that single frequency bin. The amplitude of this bin is therefore proportional to the absolute value of inner product between the $k$th class weight vector $\boldsymbol{c}^{(k)}$ and the Layer 1 output vector $\boldsymbol{d}$:

$$w_k = \mu \left| \sum_{j=1}^{Q} c_{kQ+j} d_j \right| = \mu \left| \boldsymbol{c}^{(k)} \cdot \boldsymbol{d} \right|,\ \ k = 0,1, \dots, n - 1,$$

where $\mu$ is the nonlinear conversion efficiency coefficient determined by the second-order nonlinear system. Thus, Layer 2 is decoded by reading the $n$ frequency bins located at:

$$\Omega_{w,k} = 2\Omega_0 + [K + (k + 1)Q + 1]\Omega_m,\ \ k = 0,1, \dots, n - 1.$$

For the four-class classification task demonstrated in this work, $n = 4$ and these four readout bins give the four class scores. The class corresponding to the largest score is selected as the predicted output label.

**Supplementary Table 1 | Encoding settings used in optical implementation**

Note: The definitions of $\Omega_0$, $\Omega_m$, and $K$ follow Equations (S1), (S2) and (S4).

| | $\Omega_0/2\pi$ (Hz) | $\Omega_m/2\pi$ (Hz) | $K$ | $V_{pp}$ (V) |
|---|---|---|---|---|
| Layer 1 | $1\times10^6$ | 100 | - | 2.5 |
| Layer 2 | $5\times10^6$ | 100 | 45000 | 5.0 |

**Supplementary Table 2 | Encoding settings used in electronic implementation**

Note: The definitions of $\Omega_0$, $\Omega_m$, and $K$ follow Equations (S1), (S2) and (S4).

| | $\Omega_0/2\pi$ (Hz) | $\Omega_m/2\pi$ (Hz) | $K$ | $V_{pp}$ (V) |
|---|---|---|---|---|
| Layer 1 | $10\times10^6$ | 500 | - | 2.0 |
| Layer 2 | $15\times10^6$ | 500 | 26000 | 2.5 |

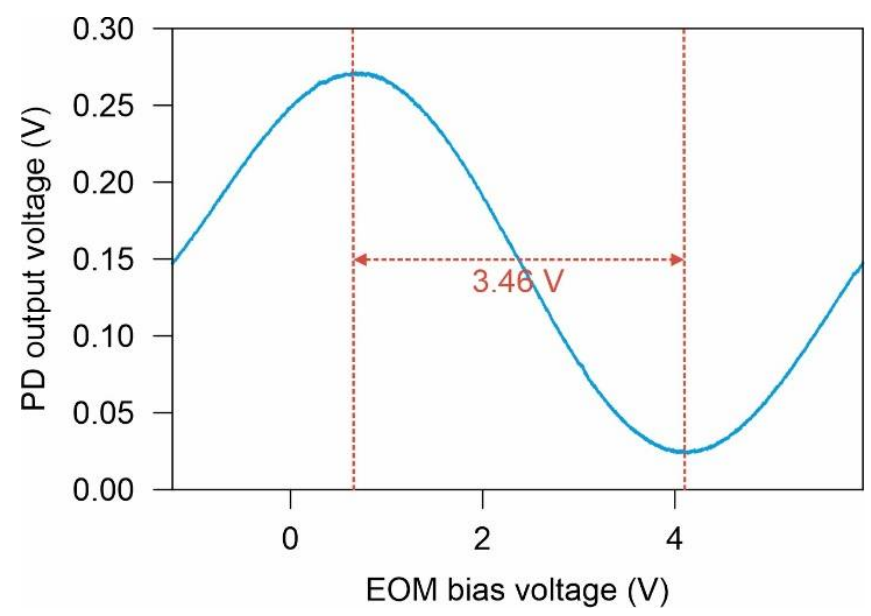

**Supplementary Figure 1 | Transfer function characterization and half-wave voltage ($V_\pi$) measurement of the electro-optic intensity modulator (EOIM).** Measured photodetector (PD) output voltage as a function of the DC bias voltage applied to the EOIM. The extracted half-wave voltage ($V_\pi$) is 3.46 V, defined as the voltage difference between the transmission peak and adjacent transmission valley.

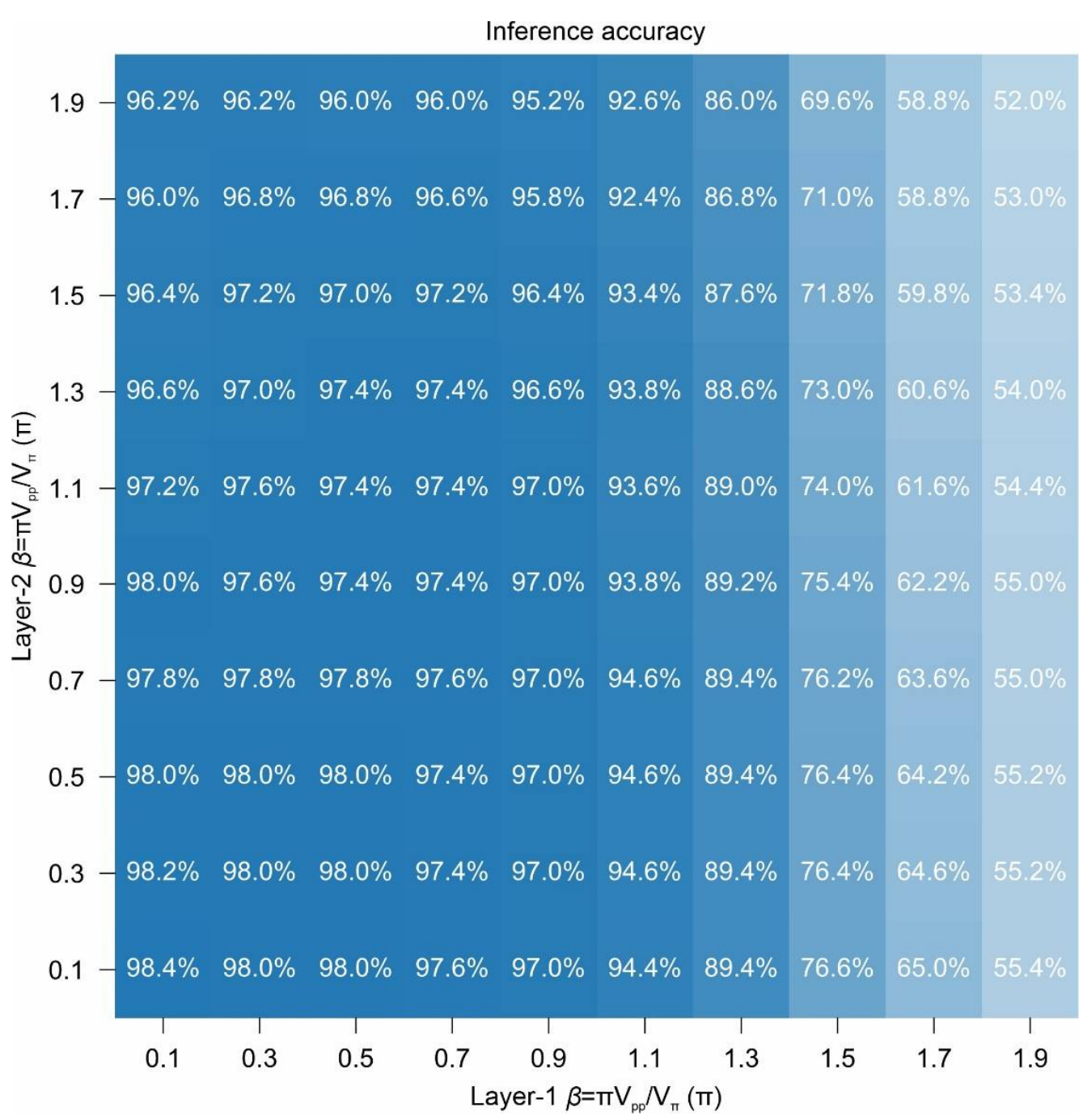

**Supplementary Figure 2 | Dependence of inference accuracy on the peak-to-peak voltage ($V_{pp}$) of the electro-optic intensity modulator (EOIM) driving signal.** The EOIM is biased at the quadrature point, and the experimental noise is excluded in these numerical calculations. The inference accuracy shows a significantly stronger dependence on the value of Layer-1 $V_{pp}$ than on the Layer-2 $V_{pp}$.

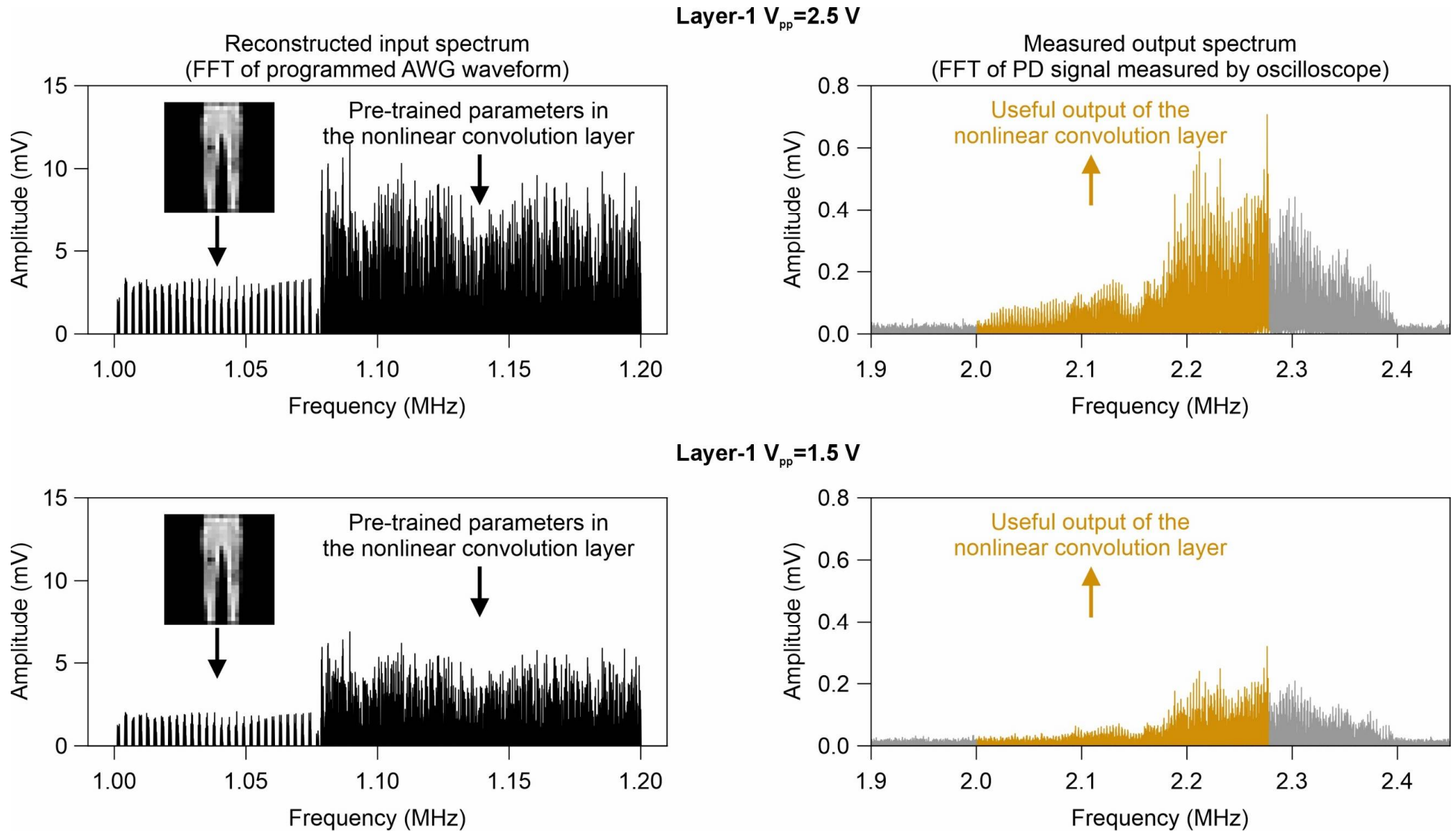

**Supplementary Figure 3 | Effect of the peak-to-peak voltage ($V_{pp}$) of electro-optic intensity modulator (EOIM) driving signal on the spectral signal-to-noise ratio (SNR) of the Layer 1 output.** Reconstructed input spectra (left) and measured second-order output spectra (right) of Layer 1. The $V_{pp}$ of EOIM driving signal are 2.5 V (top) and 1.5 V (bottom), respectively. The same input image and encoding settings are used in both measurements, except for the $V_{pp}$. Reducing the $V_{pp}$ of EOIM driving signal decreases the amplitudes of the encoded input frequency bins and the generated second-order output frequency bins, resulting in a visibly reduced spectral SNR in the measured output spectrum. The orange regions indicate the useful Layer-1 output for subsequent computation.

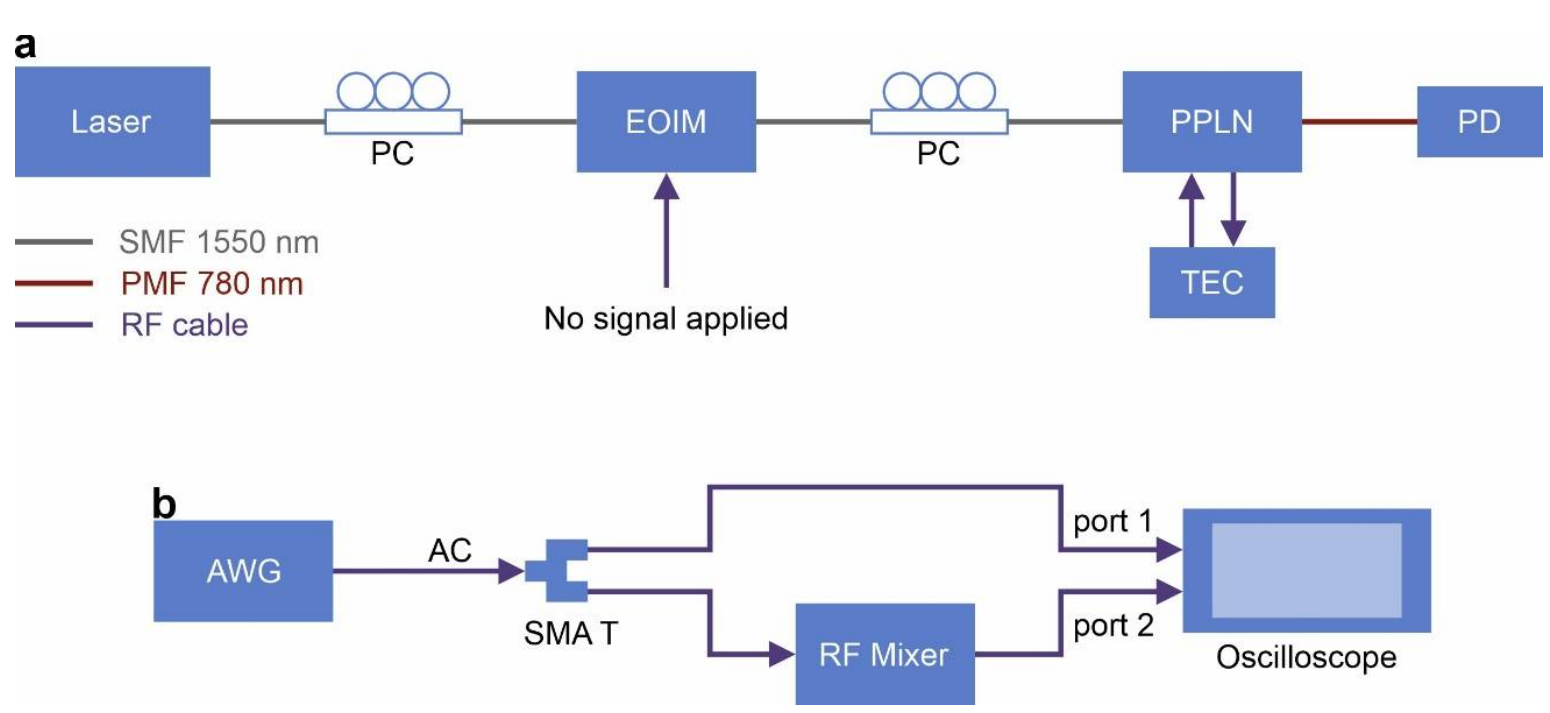

**Supplementary Figure 4 | Experimental setups for measuring nonlinear conversion efficiency. a.** Setup for the PPLN module. No DC or AC signal is applied to the EOIM. **b.** Setup for the RF mixer. Port 1 of the oscilloscope measures the reference input signal provided to the mixer via the SMA T connector, and Port 2 measures the mixer output.